\documentclass[twocolumn,linenumbers,longauthor]{emulateapj}
\usepackage{amsmath,amssymb,latexsym}
\usepackage{natbib}
\usepackage{graphicx}

\newcommand{\MJ}{M_{\rm J}}
\newcommand{\RJ}{R_{\rm J}}

\newcommand{\vecrp}{\vec{r}\,'}
\newcommand{\Ree}{\Re{}e}
\newcommand{\Imm}{\Im{}m}
\newcommand{\grad}{^{\circ}}

\received{2018 November 13}
\accepted{2019 March 1}
\shorttitle{Static tides on Jupiter}
\shortauthors{N.~Nettelmann}

\begin{document} 

\title{Tesseral harmonics of Jupiter from static tidal response}

\author{N.~Nettelmann}
\affiliation{Universit\"at Rostock, Institut f\"ur Physik, 18051 Rostock, Germany}

\begin{abstract}
%%% Context %%%
The Juno Orbiter is measuring the three-dimensional gravity field perturbation of Jupiter induced by its 
rapid rotation, zonal flows, and tidal response to its major natural satellites. 
%%% Aims %%%
This paper aims to provide the contributions to the tesseral harmonics coefficients $C_{nm}$, $S_{nm}$, 
and the Love numbers $k_{nm}$ to be expected from static tidal response in the gravity field 
of rotating Jupiter.
%%% Methods %%%
For that purpose, we apply the method of Concentric Maclaurin Ellipsoids (CMS). 
As we are interested  in the variation of the tidal potential with the
longitudes of the moons, we take into account the simultaneous presence of the satellites Io, Europa, and 
Ganymede. We assume co-planar, circular orbits with normals parallel to Jupiter's spin axis. 
The planet-centered longitude of Io in the three-moon case is arbitrarily assumed $\varphi$ = 0.
%%% Results and Conclusions%%%
Under these assumptions we find maximum amplitudes and fluctuations of $3.5\times 10^{-8} \pm 15\%$ for $C_{22}$.
For the Love numbers, largest variation of 10\% to 20\% is seen in $k_{42}$ and $k_{62}$, whereas the values 
$k_{2}$, $k_{33}$, and $k_{44}$ fall into narrow ranges of 0.1\% uncertainty or less. In particular, 
we find $k_2=k_{2,\,\rm Io}(1 \pm 0.02\%)$ where $k_{2,\,\rm Io}=0.5897$ is the static tidal response 
to lone Io. Our obtained gravity field perturbation leads to a maximum equatorial shape deformation of up to 28~m. 
We suggest that should Juno measurements of the $k_{nm}$ deviate from those values, it may be due to
dynamic or dissipative effects on Jupiter's tidal response. 
Finally, an analytic expression is provided to calculate the tesseral harmonics contribution from static tidal 
response for any configuration of the satellites.
\end{abstract}

\keywords{planets and satellites: individual: (Jupiter)}

%%%%%%%%%%%%%%%%%%%%%%
\section{Introduction}

Gravity field observations offer a window into planetary interiors, atmospheres, and the workings of tides. 
Currently, the  Juno spacecraft \citep{Bolton17} is taking a comprehensive 3D map of Jupiter's 
gravity field. Data from the first few perijoves \citep{Folkner17,Iess18} allowed to infer the presence of 
deep zonal flows through analysis of the odd ($J_{2n+1}$) gravitational harmonics \citep{Kaspi18,Kong18}. 
The even ($J_{2n}$) harmonics in addition put constraints on the deep interior density 
distribution of the giant planets \citep{Debras19,Guillot18,Wahl17J,Nettelmann17,Miguel16,HMilz16,Helled11}, 
a property appreciated long before the advent of spacecraft at Jupiter \citep{Hubbard74,Zharkov73,DeMarcus58}.

%%% why tides (tesseral harmonics) are important %%%
Upcoming Juno data may also yield information  on the tesseral harmonics $C_{nm}$, $S_{nm}$, and the 
Love numbers $k_{nm}$ of Jupiter's gravity field up the fourth degree \citep{Tommei15}. 
Those are not only sensitive to atmospheric dynamics and non-axisymmetric density anomalies, 
but also to tides raised by the Galilean satellites. Tides are not only a phenomenon important to coastal 
dwellers on inhabited planets; 
they help shaping the orbital configuration of close-in exoplanets \citep{Jackson08,FerrazMello08}, 
can inflate their radius \citep{Miller09}, and dramatically accelerate their evaporation \citep{Li10}.  
Tides thus constitute an important astrophysical phenomenon also for giant planets.
 
%%% current meaurement and uncertainty in k2 %%%
Using Juno data, \citet{Iess18} provided observational estimates of Jupiter's 
low-order tesseral harmonics $C_{21}$, $S_{21}$, $C_{22}$ and $S_{22}$, suggesting these coefficients are 
consistent with being zero. Moreover, they obtained the first observational estimate of Jupiter's Love number $k_{22}$, shortly after the first $k_{22}$ measurement for a giant planet ever had been presented, in that case from Cassini observations of Saturn \citep{Lainey17}. 

%%% current state: model k2 agrees with observed k2 %%%
Both the Jupiter and Saturn observed $k_{22}$ values, often abbreviated $k_{2}$, are consistent 
with predictions from respective interior models \citep{Wahl16,Wahl17S,Lainey17}. This is good since for a 
fluid planet, the Love number $k_{2}$ is known to be a measure of central mass condensation and thus to be 
susceptible to the core mass \citep{GavZha77}; the current agreement lends confidence to our understanding 
of the planets' mass distribution, which is derived primarily from analysis of the low-order $J_{2n}$.

%%% tight k2 range from interior models %%%
On the other hand, theoretically predicted $k_{2}$ value ranges can be strikingly tight. For Saturn, 
\citet{Kramm11} found that if the internal mass distribution is already constrained by $J_2$ and $J_4$, 
then $k_2$ will as well adopt a certain value. For Jupiter, \citet{Wahl16} quantified the relative variation 
in $k_{2}$ to be only $\sim 0.03\%$ if their adiabatic interior models were allowed to differ in  
$J_4$ value by up to 2\%, a range spanning what is now $10^4$ times the current Juno $1\sigma$ observational 
uncertainty, though \citet{Ni18} obtained a larger variation in $k_{2}$ of 2\% under the assumption of
polynomial density profiles but tightly constrained $J_4$ value. These current $k_{2}$ predictions 
from interior models are based on assumed static tidal response. 

Improved longitudinal coverage of Jupiter during the further course of the Juno mission is expected to 
significantly decrease the observational uncertainty in $k_{2}$ from its current $3\sigma$ value of 10\%,  
down to $\sim 10^{-3}$ \citep{Serra16}, though that estimate was based on the originally planned 11-day orbit
and not on the actual 53-orbit. 

%%% Temporal variation in Cnm and Snm %%% F
The strong correlation between $k_{2}$ and $J_2$ seen in the planet interior models together with an increasingly 
accurate observational grip on $k_{2}$ poses an interesting situation, as any deviation between measured and 
predicted values from a static theory is then supposedly due to dynamic effects --or due to a resilient gap 
in our understanding of Jupiter's internal structure.

%%% Dynamic tides new
Dynamic tides on a planet can occur as a result of planetary oscillations. 
Acoustic oscillations may already have been detected on Jupiter \citep{Gaulme11}, but their driving
mechanisms remain a puzzle \citep{DederickJack17,Dederick17}.

%%% knm not always entirely insensitive %%%
In contrast to the insensitivity of Jupiter's $k_{2}$ to different internal structure models, strong variation 
is seen in dependence on the single satellite considered (\citealp[][hereafter WHM16]{Wahl16}). This applies 
to some $k_{nm}$ and much more so to the $C_{nm}$. However, when Juno is flying close to Jupiter, it sees a 
snapshot of the tides raised by the temporal configuration of all the major satellites, not just from a
single one.

%%% this work %%%
The aim of this Paper is to predict the values and range of variation of the Love numbers $k_{nm}$ and 
the corresponding magnitude of the tidal potential in form of tesseral harmonics $C_{nm}$, and $S_{nm}$
under the assumption of a static tidal response. Dynamic effects are beyond the scope of the present work.  
In this Paper, the use of the harmonic coefficients $C_{nm}$, $S_{nm}$ is different from that adopted in planetary
geodesy. Here $C_{nm}$, $S_{nm}$ are quantities proportional to the instantaneous tidal potential
due to a given moon. 
To compute the tidal field we employ the CMS method \citep{Hubbard13,Wahl17S}. 
Methods are presented in Section \ref{sec:method} and in Appendix \ref{sec:apx}. 
As this paper aims at assessing the variation of the response at the orbital frequencies of Io, Europa and Ganymede, 
the underlying interior model used in this work is the same for all Love number computations and is described in Section \ref{sec:Jmodel}.
In Section \ref{sec:res_lone} we consider the case of lone but different Galilean satellites at 
planet-fixed longitude $\varphi=0$, while in  Section \ref{sec:res_Io} we let Io's longitude vary. 
In section \ref{sec:res_three} we compute the range of variation due to the simultaneous presence of 
the three closest satellites. Conclusions are given in Section \ref{sec:conclusions}.

%%%%%%%%%%%%%%%%%%%%%%%%%%%%%%%%%%%%%%%%%%%
\section{Methods}\label{sec:method}

In this work we compute the tesseral $(0<m\leq n)$ harmonics $C_{nm}$ and $S_{nm}$ (e.g.;~\citealp{Kozai61}), 
\begin{equation}\label{eq:Cnm}
	\MJ\RJ^n \,C_{nm} = 2 \frac{(n-m)!}{(n+m)!} \:\int_{r'<R}\!\!\!\! d^3r'\,\rho\: r'^{\,n} \cos m\varphi' \,
	P_n^m(\mu')\:,
\end{equation}
\begin{equation}\label{eq:Snm}
	\MJ\RJ^n \,S_{nm} = 2 \frac{(n-m)!}{(n+m)!} \: \int_{r'<R}\!\!\!\! d^3r'\,\rho\: r'^{\,n} \sin m\varphi' \,
	P_n^m(\mu')
\end{equation}
of the planetary gravity field of rigidly rotating Jupiter of mass $\MJ$ and equatorial radius $\RJ=71,492$~km. 
The integrals in Eqs.~(\ref{eq:Cnm},\ref{eq:Snm}) are taken over the entire planetary volume enclosed by the 
equipotential surface $R(\varphi,\vartheta)$, $\rho$ denotes mass density, $P_n^m$ are the associated Legendre 
polynomials, and $\mu=\cos\vartheta$.  By definition, the static fluid Love numbers $k_{nm}$ are the linear 
response coefficients of the components $V_{nm}$ of the distorted planetary gravity field $V$ in response to 
the components $W_{nm}$ of the tide-raising gravitational potential $W$, thus $V_{nm}=k_{nm} W_{nm}$. 
Expressions for the components $V_{nm}$ and $W_{nm}$ are derived in Appendix \ref{sec:apx}.

We assume that the satellites $S_{i},\:i=1,2\ldots,N_{\rm  sat}$ are located in the planetary equatorial 
plane ($\mu=\mu_{S_i}=0$). The influence of any deviation from this symmetric case on the tidal field is 
supposedly small, due to the small orbital inclinations of the satellites of less than $0.5^{\circ}$ 
\citep{Peale99}. Non-zero inclinations would lead to non-zero values of harmonics such as $C_{21}$ that 
otherwise must vanish if northern and southern hemispheres were fully symmetric and the rotation axis aligned 
with the principal axis of inertia. Since the satellites act as independent point sources on the planet, the 
tide-raising potential can be written as linear superposition of the single-satellite contributions,
\begin{equation}
	W(\vec{r}) = \sum_{i=1}^{N_{\rm sat}} GM_{S_i}/|\vec{r}-\vec{r}_{S_i}|,
\end{equation}  
where $M_{S_i}$ and $r_{S_i}$ are the satellite's mass and orbital distance measured from the center of the 
planet at $\vec{r}=0$. We also assume that planetary tidal response happens instantaneously. Dynamic and 
non-linear effects are neglected. For multiple satellites $\{S_i\}$ we find 
\begin{equation}\label{eq:knm_multi1}
	k_{nm} = -\frac{3}{2}\,\frac{(n+m)!}{(n-m)!}\Bigg[  
		\frac{ C_{nm}\sum c_{nmi}	+ S_{nm}\sum s_{nmi}}{ (\sum c_{nmi})^2 + (\sum s_{nmi})^2}\Bigg]
\end{equation}
where 
\[
	\sum c_{nmi}:=\sum_{i=1}^{\rm N_{sat}} q_{Si}\Big(\frac{R_{eq}}{r_{S_i}}\Big)^{n-2}\:P_n^m(\mu_{S_i})
	\:\cos m\varphi_{S_i}\:,
\]
\[
	\sum s_{nmi}:=\sum_{i=1}^{\rm N_{sat}} q_{Si}\Big(\frac{R_{eq}}{r_{S_i}}\Big)^{n-2}\:P_n^m(\mu_{S_i})
	\:\sin m\varphi_{S_i}\:,
\]
and $q_{Si} = -3(M_{S_i}/M_p)(R_{\rm eq}/r_{S_i})^3$ denotes the tidal forcing of satellite $S_i$ and $\varphi_{S_i}$
its orbital longitude with respect to a planet-centered reference frame corotating with Jupiter. Since we 
neglect dynamic effects and dissipation, Eq.~(\ref{eq:knm_multi1}) reduces to
(see Appendix \ref{sec:apx_knm})
\begin{equation}\label{eq:knm_multi2}
	k_{nm} = -\frac{3}{2}\,\frac{(n+m)!}{(n-m)!}\Bigg[  
		\frac{ C_{nm} }{ \sum c_{nmi} }\Bigg]\:,
\end{equation}

%%% numerics %%%
which agrees with Equation (41) in \citet{Wahl17S} for a single moon ($N_{Sat}=1)$.
To numerically compute the $C_{nm}$ and $S_{nm}$ in a selfconsistent manner with the tri-axial shape of the 
rotationally and tidally deformed planet we apply the CMS method described in \citet{Wahl17S} 
with two simplifications. First, we compute integrals over  colatitude $\vartheta$ and the azimuthal angle 
$\varphi$ using Legendre-Gauss Quadrature in both cases instead of recasting integrals over $\varphi$ to 
Gauss-Chebychev integration. Second, we do not apply the correction to the shape due to the center of mass shift, 
leading to non-zero $C_{11}$ values $\sim 10^{-12}$, whereas $C_{11}$ should be zero if the barycenter is 
the origin of the reference frame.  As this inherent inaccuracy might affect other moments as well, we only 
display moments for which $C_{nm}\gg 10^{-12}$, except for $n=m=4$.

%%% Jupiter model %%%%%%%%%%%%%%%%%%%%%%%%%%
\subsection{Jupiter model}\label{sec:Jmodel}

For the interior density distribution $\rho(r)$ we take the 2D model J17-3b of \citet{Nettelmann17} as an 
initial guess.  This is a model for Jupiter rotating rigidly at a rate corresponding to a rotational forcing
$q_{rot}=0.089195486$. That model gives a reasonable match to the low-order gravitational harmonics 
measured by Juno   \citep{Iess18}. The deviations amount to $20\times$, $30\times$, and $4\times$ the observational 
$1\sigma$ error bars in $J_2$, $J_4$, $J_6$, respectively; however, their relative deviations are only 
$6\times10^{-6}$, $7\times10^{-5}$, and $5\times10^{-3}$. These discrepancies are irrelevant to the further
analysis carried out in this paper. For the computations in 3D that include the tidal distortion, we here use a 
smaller number of radial grid points, $N_{\rm CMS}=256$ instead of 1000. This yields an up to 10 times 
stronger deviation from the observed low-order $J_{2n}$ values. 

Obtaining converged gravitational harmonics $J_n$ in the 2D case (rotation only) requires about 40 iterations 
between the potential and the shape. Starting with the given 2D density distribution we do this in 3D  
and obtain  the Love numbers $k_{nm}$ converged  within $10^{-6}$. The 3D calculations of this work use
$N_{\rm POL}=22$ latitudinal grid points, $N_{\rm AZ}=24$ azimuthal grid points, and extend the order of 
expansion to $2n\leq 20$ and $m\leq 8$%
\footnote{For this choice of numbers we find relative convergence in $J_{2n=10}$ within $10^{-5}$ under small variation
in  $N_{\rm POL}$; the convergence of the $k_{nm}$ was tested against the analytic result $k_n=3/2(n-1)$ for 
a weakly rotating ($q_{\rm rot}\ll 1$) homogeneous ($N=1$) body, obtaining $10^{-5}$ variation for $n<5$ and 
$10^{-4}$ for $n<8$. Larger number of grid points would be required for accurate computation of higher degrees 
$2n>10$ and orders $m>8$.}.

%%%% Table 1 %%%%%%%

\renewcommand{\arraystretch}{0.9}
\begin{deluxetable*}{c|rr|rrr}
\tablecaption{Static response of Jupiter to single satellites\label{tab:singlemoons}}
\tablecolumns{6}
%\tablenum{}
\tablewidth{0pt}
%\rotatebox{270}{
\tablehead{
\colhead{} & \colhead{Io} & \colhead{Io-W16$^a$} & \colhead{Europa} & \colhead{Ganym.} & \colhead{Callisto}
}
\startdata
\small
$k_{2}$  	& 0.58970 	&	0.58985			& 0.58934 			& 0.58921 			& 0.58915  \\
$k_{31}$ 	& 0.24168 	&	0.19118			& 0.24154 			& 0.24143 			& 0.24148 \\
$k_{33}$ 	& 0.23944 	& 	0.23989			& 0.23928 			& 0.23920 			& 0.23883 \\
$k_{42}$ 	& 1.75678 	& 	1.75874			& 4.28117 			& 10.7244 			& 32.9489 \\
$k_{44}$ 	& 0.13529 	& 	0.13537			& 0.13403 			& 0.12854 			& -- \\
$k_{51}$ 	& 1.09441 	& 0.95088			& 2.67351 			& 6.70277 			& 20.5324 \\
$k_{53}$ 	& 0.82177 	& 0.82162			& 1.96567 			& 4.85501 			& 0.49505 \\
$k_{62}$ 	& 5.99597 	& 5.98975			& 35.9649 			& 226.794 			& 2135.49 \\\hline
$C_{22}$ 	& 3.3777 -08 		& 	--			& 4.5041 -09		& 3.4261 -09 		& 4.5710 -10 \\
$C_{31}$ 	& 2.3463 -09 		& 	--			& 1.9665 -10 		& 9.3764 -11 		& 7.1145 -12 \\
$C_{33}$ 	& $-3.8745$ -10 	& 	--			& $-3.2468$ -11	& $-1.5483$ -11 	& $-1.1727$ -12 \\
$C_{42}$ 	& $-4.8179$ -10 	& 	--			& $-6.1885$ -11	& $-4.6364$ -11 	& $-6.1437$ -12 \\
$C_{44}$ 	& 4.6380 -12 		& 	--			& 2.4219 -13		& 6.9463 -14 		& -- \\
$C_{51}$ 	& $-1.5261$ -10 	& 	--			& $-1.2351$ -11	& $-5.8063$ -12 	& $-4.3614$ -13 \\
\enddata
\tablenotetext{a}{Label Io-W16 denotes the results by \citet{Wahl16} for their Jupiter model DFT-MD 7.15 ($J_4$).} 
\tablecomments{All $S_{nm}$ are zero because the satellites are placed at orbital longitude zero 
in the planet-fixed reference frame.}
\end{deluxetable*}

%%%%%%%%%%%%%%%%%%%%

As an upper bound estimate for the uncertainties in the tesseral harmonics, we compare in Table \ref{tab:singlemoons}
our result for the Jupiter-Io system to the result of WHM16 for their preferred Jupiter model, 
which matches $J_2$ precisely but is off in  $J_4$ by a rather large amount of 0.17\%. The differences 
are generally small; they are 0.03\% in $k_{2}$,  0.19\% in $k_{33}$, 0.11\% in $k_{42}$, 0.08\% in $k_{44}$, 
and 0.1\% in $k_{62}$; only for $k_{31}$ we obtain a significant 8\% deviation. However, for the $N_{\rm CMS}=1$ 
test case, which reproduces the exact analytic result  $k_{nm}=3/2(n-1)$ in the limit of vanishing $q_{rot}$ 
and $q_{tid}$, we obtain $k_{31}(q_{rot})$, after a shallow minimum at $q_{rot}\sim 0.01$, to be a rising function 
of $q_{rot}$, while \citet{Wahl17S} obtain $k_{31}$ to be a decreasing function of $q_{rot}$ --unlike the 
other $k_{nm}$. Therefore, we attribute the difference in $k_{31}$ for the Jupiter-Io system to whatever is the 
reason for the different behavior in the constant-density case, but not to matching the low-order $J_{2n}$ of 
Jupiter not precisely. Based on the above comparison we conservatively estimate the accuracy of our $k_{nm}$, 
and of the underlying $C_{nm}$ and $S_{nm}$ calculations, to be within 0.2\% for our Jupiter models using 
$N_{\rm CMS}=256$.

%%%%%%%%%%%%%%%%%%%%%%%%%%%%
\subsection{Polytrope model}

%%% code validation: bom-rotating n=1 polytrope %%%
The non-rotating polytrope of index $n=1$ offers the possibility to estimate the accuracy of the numerical 
computation  since its static $k_{2}$ value is analytically known; it is $k_2=15/\pi^2-1.$ 
Our results differ from that value by 0.8\%, 0.23\%, 0.07\%, 0.02\%, 0.006\% for $N_{\rm CMS}=64$, 128, 256, 
512, and 1024 respectively, while \citet{Wahl17S} achieve 0.008\% deviation for only 128 CMS layers.

The difference between the latter and this work may partially be due to different spheroid partitionings;
especially the size of the outermost layer has been shown to affect the accuracy of the resulting gravitational
harmonics \citep{Debras18}. Here we use a spacing similar to the one of the Jupiter models in \citet{Nettelmann17}, 
which decreases toward center and surface and has a size ratio of 1/3 between the first and the second layer.

%%%%%%%%%%%%%%%%%%%%%%%%%%%%
\subsection{Tidal forcings}

The relative contributions of Europa, Ganymede, and Callisto to the maximum amplitude of the tide-raising 
potential $W$ depend on their tidal $q$-ratio with respect to that of satellite Io. These ratios are 0.133, 0.101, 
and 0.013, with  $q_{\rm Io}$=$-6.872\:10^{-7}$, $q_{\rm Eur}$=$-0.917\:10^{-7}$, $q_{\rm Gan}$=$ -0.698\:10^{-7}$, 
and $q_{\rm Cal}$=$-0.093\:10^{-7}$ using online data\footnote{\tiny{https://nssdc.gsfc.nasa.gov/planetary/factsheet/joviansatfact.html}}. For comparison, $q_{tid}$=$-1.687\:10^{-7}$ 
for the Earth-Moon system, and $q_{tid}$=$-0.775 \:10^{-7}$ for the Earth-Sun system.

%%%%%%%%%%%%%%%%%%
\section{Results}

%%%%%%%%%%%%%%%%%%%%%%%%%%%%%%%%%%%%%%%%%%%%%%%%%%%%%%%%%%%%%%%%%%%%%
\subsection{Single Galilean satellites at Jupiter}\label{sec:res_lone}

In this Section we investigate the resulting $k_{nm}$ and $C_{nm}$ values for the different tidal forcings from 
single Galilean satellites. As in WHM16 each satellite is placed at $\varphi_{S_i}=0$ and $\mu_{S_i}=0$ 
one after another while the contributions from other satellites are ignored. Results are presented in Table 
\ref{tab:singlemoons}.

Due to the assumed symmetry between northern and southern hemisphere in this case, all coefficients 
$k_{nm}$, $C_{nm}$, and $S_{nm}$ are zero unless $n-m = 2i$,$i=0,1,\rm etc.$; all $S_{nm}$ are zero because 
of the assumed zero phase lag and alignment of the tidal bulge with the planet--satellite connecting axis.
In the real system, finite values of these coefficients can still occur because of non-zero satellite 
orbital inclinations or,  in particular for the low-degree harmonics $C_{21}$ and $S_{21}$, because of 
rotation-axis and inertia-axis misalignment, a phenomenon well-studied on Earth (\citealp[e.g.][]{Tamisiea02}).

In agreement with WHM16 we find that $k_{2}$ emerges as the most insensitive parameter, although the variation 
due to different single satellites is non-zero; it amounts to 0.0007, or 0.12\% and is thus one order of 
magnitude larger than the variation in $k_{2}$ of 0.01\% for the different interior models of same $J_4$ value  
by WHM16. In contrast, the variation in $k_{42}$, $k_{51}$, and $k_{62}$ can be a factor of $\sim 10$ to 
$\sim 100$ larger. The difference in the $C_{nm}$ is generally larger because of the scaling with the 
different $q_{\rm tid}$ values. Callisto's induced $C_{nm}$ values are about one order of magnitude 
smaller than those of Europa and Ganymede, which is consistent with its lower tidal $q$ value. Notably, 
our $k_{44}$ value for Io agrees well with that of WHM16 although $C_{44} \sim C_{11} \sim 10^{-12}$ in 
our calculations rather than $C_{11}=0$.

%%%%%%%%%%%%%%%%%%%%%%%%%%%%%%%%%%%%%%%%%%%%%%%%%%%%%%%%%%%%%%%%
\subsection{Single Io at different longitudes}\label{sec:res_Io}

\begin{figure}[h]
\centering
\includegraphics[angle=0,width=0.48\textwidth]{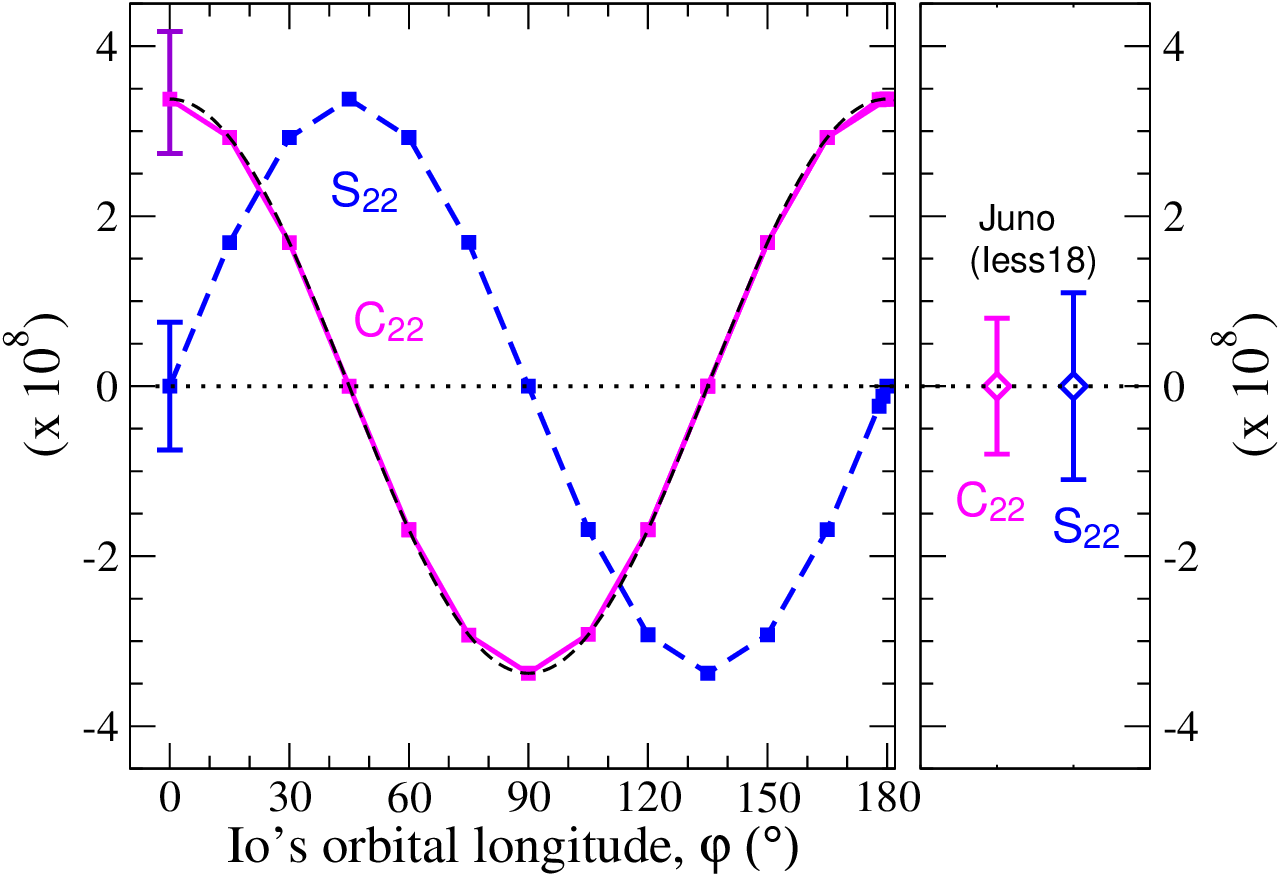}
\caption{Tesseral harmonics $C_{22}$ and $S_{22}$ of Jupiter due to assumed static tidal response to lone
Io at different orbital longitudes in the planet-fixed reference frame (left panel) and observed values
(right panel). The vertical uncertainty bars in the left panel are the fluctuations described in Section 
\ref{sec:res_three}. The black dashed curve shows the analytic result for $C_{22}(\varphi_{\rm Io})$.
\label{fig:C22S22}}
\end{figure}

In Section \ref{sec:res_lone} we assumed the satellites to be at orbital longitude $\varphi_{S_i}=0$ of the 
planet-fixed reference frame. However, this is not necessarily the case when the Juno measurements are taken. 
In Figure \ref{fig:C22S22} we plot the $C_{22}$ and $S_{22}$ values of lone Io when it is assumed to by at any 
longitude between $0$ and $\pi$. 
%%% Figure %%%%%%%%
%%%%%%%%%%%%%%%%%%%s

Placing Io at planet-centered longitude $\varphi=0$ (Section \ref{sec:res_lone}) or $\pi$ leads to the maximum possible value of $C_{22}$ from Io's static tide. In contrast, the computed $k_{nm}$ are independent of the position 
of a single moon. Comparison with the measured values \citep{Iess18} suggests that those measurements were taken 
at intermediate longitudes.

This maximum value of $C_{22}$ can still fluctuate due to the different orbital locations of the other moons 
whenever Io crosses Jupiter's meridian. These fluctuation in $C_{22}$ and $S_{22}$ are indicated by the vertical uncertainty bars in Figure \ref{fig:C22S22}. They will be addressed in Section \ref{sec:res_three}.

The general shape of the curves $C_{22}(\varphi_{\rm Io})$ and $S_{22}(\varphi_{\rm Io})$ suggests that
$C_{22}(\varphi_{\rm Io}) = C_{22}(0)\times \cos(2\varphi_{\rm Io})$ and
$S_{22}(\varphi_{\rm Io}) = C_{22}(0)\times \sin(2\varphi_{\rm Io})$. The $S_{nm}$ do not contain additional
information because we request to have no dissipiation in the system, thus they can be expressed through 
the $C_{nm}$.

%%%%%%%%%%%%%%%%%%%%%%%%%%%%%%%%%%%%%%%%%%%%%%%%%%%%%%%%%%
\subsection{Three satellites at Jupiter}\label{sec:res_three}

In this Section we investigate the range of variation in the $C_{nm}$, $S_{nm}$, and $k_{nm}$ due to Jupiter's 
static tidal response to the three simultaneously present satellites Io (at $\varphi_{\rm Io}=0$), Europa, 
and Ganymede. Callisto is neglected due to its minor tidal influence on Jupiter. 
However, Europa and Ganymede cannot be anywhere when Io is at $\varphi=0\grad$; their orbital longitudes 
are related to each other and to Io through the 4:2:1 Laplace resonance.
We idealize the Laplace resonance configurations of the three inner satellites by assuming perfect mean motion 
resonance, where the orbital mean motions would behave as $n_{\rm Io}$ : $n_{\rm Eur}$ : $n_{\rm Gan} =$ 4:2:1, 
and the satellites are on perfect circular orbits, so that $\Omega_{i}=n_i$.

%%% Figure %%%%%%%%
\begin{figure*}[t]
\centering
\includegraphics[angle=0,width=0.92\textwidth]{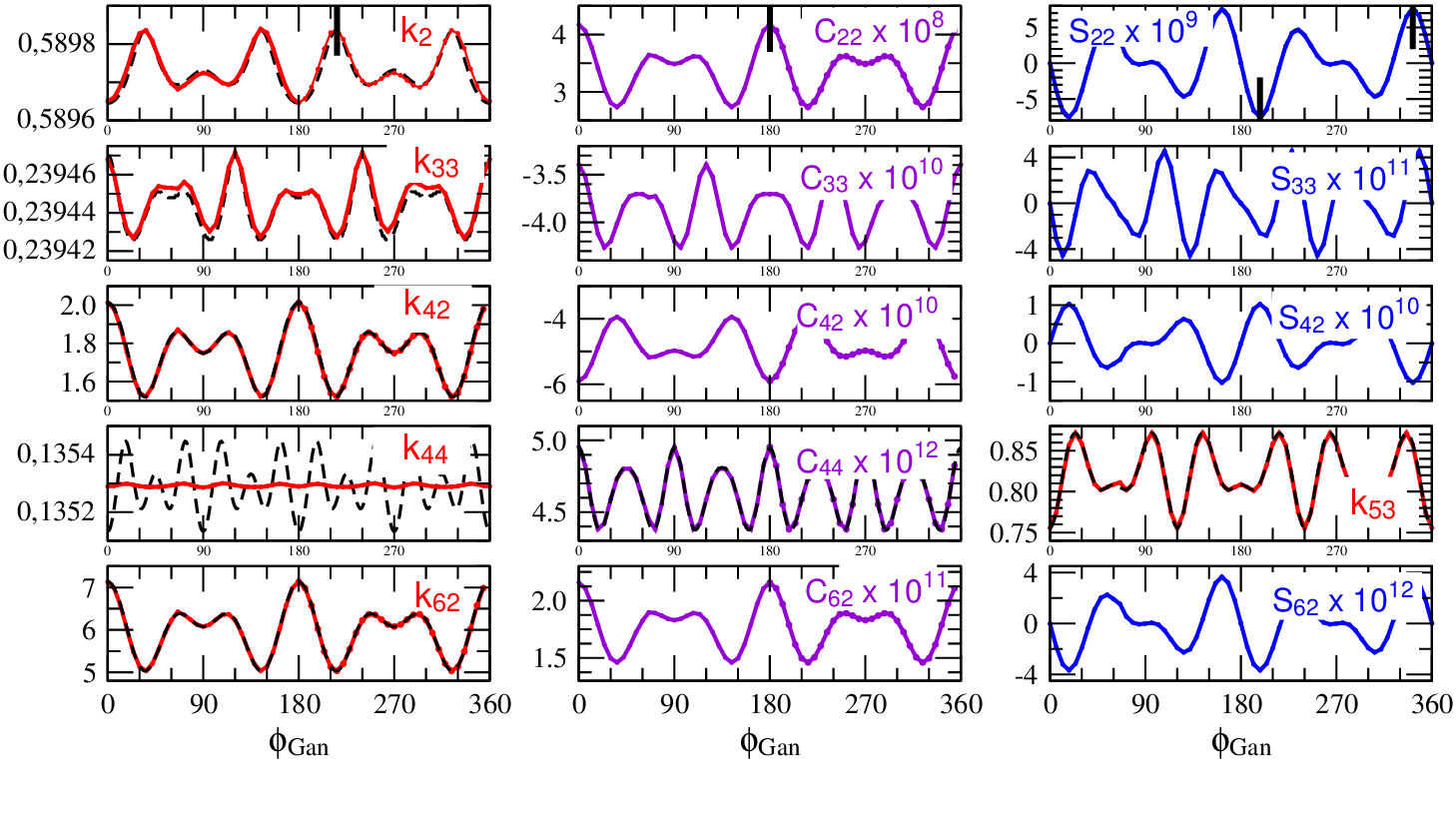}
\caption{Tesseral harmonics of Jupiter and their fluctuation due to different orbital positions of Europa 
and Ganymede as labeled by the support coordinate $\phi_{\rm Gan}$ (see text for explanation), while Io resides
at orbital longitude $\varphi=0$ in the planet-fixed reference frame. Vertical bars in the upper row of panels 
mark the configurations for which we present the shape deformation in Fig.~\ref{fig:shapes}. 
Overplotted black dashed curves in some of the panels show the analytic expressions using 
Eqs.~(\ref{eq:knm_multi1})--(\ref{eq:Snm_ana}).}
\label{fig:knmCnmSnm}
\end{figure*}
%%%%%%%%%%%%%%%%%%%s

In the real system, the eccentricities are, 
partially due to perturbation by Callisto and partially due to the Keplerian orbits, maintained at small 
but non-zero value of $e_{\rm Gan}=0.001<e_{\rm Io}< e_{\rm Eur}=0.01$ \citep{Peale99} so that $\Omega_i$ 
unequal $n_i$. Still, the two-body 2:1 mean motion resonances of Io and Europa and Europa and Ganymede hold 
approximately and the three-body mean motion resonance is well satisfied  \citep{Sinclair75}. 
Technically, we consider one orbit of Ganymede and parametrize it by the variable $\phi_{\rm Gan}$.
Note that $\phi_{\rm Gan}$ does not refer to any reference system. It it just used to conveniently 
calculate the possible relative orbital distances of the three satellites. For example, $\phi_{\rm Gan}=\pi/4$ 
implies that Ganymede lags behind Io, which then has $\phi_{\rm Io}=\pi+4\times\pi/4$, by 
$\Delta\phi=2\pi-\pi/4=315\grad$ and thus would be at planet-centered longitude $\varphi=45\grad$, while Europa 
has $\phi_{\rm Eu}=2\times\pi/4$ and thus lags behind Io by  $\Delta\phi=2\pi - \pi/2$ and thus would be at 
planet-fixed longitude $\varphi=90\grad$.

%%% Results from Figure 2: k2 %%%
Figure \ref{fig:knmCnmSnm} shows a selection of the resulting tesseral harmonics and Love numbers.
We find the amplitude of the variation in $k_{2}$ due to the different positions of the inner three moons 
to be 0.0001, or 0.017\%. This is a factor of seven smaller than in the artificial single satellites setup 
of Section \ref{sec:res_lone}, but a factor of 1.7 larger than the variation due to different Jupiter models 
of same $J_2$ and $J_4$ value according to WHM16. Since we find $k_2 = 0.5896$--0.5898 and 
$k_{2,\,\rm Io}=0.5897$ (Table~\ref{tab:singlemoons}), we conclude that Jupiter's static $k_{2}$ value is 
$k_2=k_{2,\,\rm Io}(1 \pm 0.02\%)$. 
%%% Results from Figure 2: k31, k33 %%%
%As $k_{33}$ falls within the narrow range of 0.2392--0.2399 for all three single satellites 
%(Table \ref{tab:singlemoons}), it is not surprising that we also obtain $k_{33}=k_{33,\,\rm Io}(1\pm 0.01\%)$.
%In contrast, the $k_{31}$ value of the three-satellites system of $\sim 0.281$ turns out to be outside the range
%of 0.2414--0.2417 suggested by the single satellite contributions, whereas the $k_{31}$ values of WHM16 cluster even
%farther away near 0.194. At present this behavior remains unexplained. 
We also obtain $k_{33}=k_{33,\,\rm Io}(1\pm 0.01\%)$.

%%% Results from Figure 2: k42, k44, k62 %%%
Particularly large variations are seen in $k_{42}$ and $k_{62}$, though the variation is a factor of several 
lower compared to the single satellite contributions of Section \ref{sec:res_lone}. We find
$k_{42}=k_{42,\,\rm Io}(1\pm 10\%)$ and %$k_{44}=k_{44,\,\rm Io}(1\pm 0.007\%)$, 
and $k_{62}=k_{62,\,\rm Io}(1\pm 20\%)$.

In contrast, the $C_{nm}$ values generally show large variations in dependence on the satellite positions
of $\pm 20\%$ ($C_{42}$), $\pm 15\%$ ($C_{22}$, $C_{62}$), and $\pm 10\%$ ($C_{31}$, $C_{33}$), 
%$\pm 6$ to 8\% ($C_{44}$, $C_{51}$), 
while the $S_{nm}$ can be raised to non-zero values. 
%of the order of $10^{-9}$ ($S_{22}$), $10^{-10}$ ($S_{31}$, $S_{42}$), or smaller. 
These results lead us to conclude that taking into account the actual orbital positions of Io, Europa, and
Ganymede is important for an observational determination of the static contribution to Jupiter's  
tidal field below the 10\% accuracy level since naturally, the gravity field fluctuates in the presence of 
the natural satellites.

The behavior of the $k_{nm}$, $C_{nm}$, and $S_{nm}$ displayed in Figure \ref{fig:knmCnmSnm} is described 
by the analytic expression (\ref{eq:knm_multi1}) or, equivalently (\ref{eq:knm_multi2}) with 
\begin{eqnarray}
\label{eq:Cnm_ana}
	C_{nm} &=& \sum_{i=1}^{N_{\rm sat}} C_{nm}^{(\rm S_i)}(0)\times\cos(m\varphi_{\rm S_i}) \\
\label{eq:Snm_ana}
	S_{nm} &=& \sum_{i=1}^{N_{\rm sat}} C_{nm}^{(\rm S_i)}(0)\times\sin(m\varphi_{\rm S_i})\:.
\end{eqnarray}
Note that the only input parameters are the $C_{nm}^{(\rm S_i)}(0)$ of each lone satellite at
$\varphi_{\rm S_i}=0$ and their actual longitudes $\varphi_{\rm S_i}$. The $S_{nm}$ can be derived from the 
$C_{nm}$ according to Eq.~(\ref{eq:Snm_ana}) if alignement of the tidal bulges toward the satellites is assumed.
These relations can be used to predict the static tidal field contributions to the tesseral harmonics 
and Love numbers for any orbital configuration of multiple satellites, for instance those occuring at the times
of the Juno measurements; angles $\varphi_{S_i}$ refer to the planet-fixed coordinate system.
Comparison to the numerical results of Figure \ref{fig:knmCnmSnm} shows good agreement except for 
$k_{33}$ and $k_{44}$. The fluctuation in these coefficients is small and may be at the limit of our 
numerical resolution in the multiple-satellites simulation.

%%% Figure %%%%%%%%
\begin{figure*}[t]
\centering
\includegraphics[angle=0,width=0.84\textwidth]{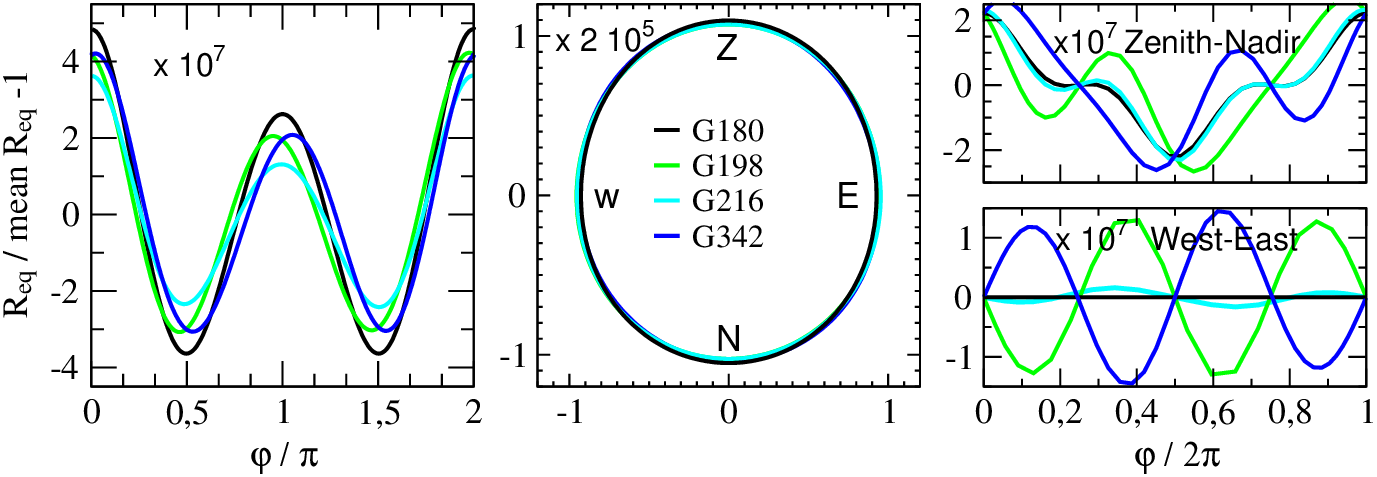}
\caption{Jupiter's shape $R_{\rm eq}(\varphi)$ for four of the configurations of the three satellites 
Io, Europa, and Ganymede for which the signature in Jupiter's gravity field is largest;
G180 (black): $\phi_{\rm Gan}-\phi_{\rm Io}=0$, $\phi_{\rm Eur}-\phi_{\rm Io}=180^{\circ}$,
G198 (green): $\phi_{\rm Gan}-\phi_{\rm Io}=306^{\circ}$, $\phi_{\rm Eur}-\phi_{\rm Io}=144^{\circ}$,
G216 (cyan): $\phi_{\rm Gan}-\phi_{\rm Io}=252^{\circ}$, $\phi_{\rm Eur}-\phi_{\rm Io}=108^{\circ}$,  
G342 (blue): $\phi_{\rm Gan}-\phi_{\rm Io}=234^{\circ}$, $\phi_{\rm Eur}-\phi_{\rm Io}=216^{\circ}$.
Jupiter's longitude $\varphi=0$ is chosen to be the direction of Io (zenith). 
Left: radius deviation with respect to mean equatorial radius with deviation exaggerated by a factor of $10^7$; 
middle: same as left panel but plotted in polar coordinates and deviation is scaled by a factor of $2\times 10^5$; 
bottom right: relative radius difference between western ($\varphi=90^{\circ}$) and eastern hemisphere;  
top right: zenith ($\varphi=0^{\circ}$) and nadir directions, difference is scaled by a factor of $10^7$. 
The static tidal distortion is of the order of 1--10~m.}
\label{fig:shapes}
\end{figure*}
%%%%%%%%%%%%%%%%%%%s

%%%%%%%%%%%%%%%%%%%%%%%%%%%%%%
\subsection{Equatorial Shape deformation}

%%% Results from Figure 3: shape %%%
In Figure \ref{fig:shapes} we plot Jupiter's equatorial shape $R_{\rm eq}(\varphi)$ for four of the satellite
positions for which the deformation is largest. Those are the configurations 
G180 ($\phi_{\rm Gan}=180^{\circ}$), where $C_{22}$ is at maximum,
G198 ($\phi_{\rm Gan}=198^{\circ}$), where $S_{22}$ is at minimum,
G216 ($\phi_{\rm Gan}=216^{\circ}$), where $k_{2}$ is at maximum, and   
G342 ($\phi_{\rm Gan}=342^{\circ}$), where $S_{22}$ is at maximum. 
As in Section \ref{sec:res_three} Jupiter's longitude $\varphi=0$ is chosen to point to the direction 
of Io (zenith), and west is defined to be at $\varphi=90^{\circ}$.

With a mean equatorial radius of $\bar{R}_{\rm eq}\approx R_J$, the shape deformation $R_{\rm eq}-\bar{R}_{\rm eq}$ 
in the equatorial plane reaches an amplitude of up to $4\times10^{-7}\times 7.1492\times 10^7\:$m $\approx 28$~m.
One may argue that this computed tidal deformation refers to a center of mass of Jupiter fixed in space and thus overestimate the true deformation. However, for the G180 configuration the center of mass shift would only occur
along the large axis. As the change in equilibrium shape along the short axis ($\varphi=90^{\circ},\:270^{\circ}$) 
is of similar magnitude,  we conclude that our results for the shape deformation are robust against the neglected 
center of mass shift.
Figure \ref{fig:shapes} furthermore illustrates that when the satellites are mis-aligned, the $S_{nm}$ are raised 
and a small east-west asymmetry of the order $10^{-8}\:\RJ$ (G216, G342) to $10^{-7}\:\RJ$ (G198)  can occur. 
The maximum zenith-nadir equilibrium asymmetry is a little larger due to the predominant influence of Io at 
$\varphi=0$, and may reach up to $2\times10^{-7}\times\RJ = 14\:$m.

%%%%%%%%%%%%%%%%%%%%%%%%%%%%%%%%%%%%%%%%%%
\section{Discussion and Conclusions}\label{sec:conclusions}

We have calculated the 3D gravity field of rotationally and tidally distorted Jupiter using the CMS method of
\citet{Hubbard13} and \citet{Wahl17S} and the internal density distribution of the interior model J17-3b of
\citet{Nettelmann17}. We presented Love number $k_{nm}$ and tesseral harmonics $C_{nm}$, $S_{nm}$ values for 
three setups of the Galilean satellites: single satellites at $\varphi=0$ (Section 3.1, Table 
\ref{tab:singlemoons}), Io alone at different longitudes $\varphi$ (Section 3.2, Figure~2), or Io-Europa-Ganymede 
in 4:2:1  mean motion resonance (Section 3.3, Figure 3). We provided an analytic formula to calculate these
coefficients for any orbital configuration (Eqs.~\ref{eq:knm_multi1}, \ref{eq:Cnm_ana}, \ref{eq:Snm_ana}).  

Assuming single satellites, we confirm the finding of WHM16 that the static tidal 
response of a fluid planet is not solely determined by the internal mass distribution, but is similarly or 
even more sensitive to the tide-raising perturber's mass and orbital parameters.

For the three-satellites case and $n$=$m$=2 or 3 we obtain $k_{nm}=k_{nm,\,\rm Io}(1\pm 0.02\%)$. 
Particularly large fluctuation of respectively 10\% and 20\% is seen in $k_{42}$ and $k_{62}$.
Although these values refer an assumed orbital longitude of Io $\varphi_{\rm Io}=0$, they apply to
any orbital longitude of Io, since the love numbers depend on the relative positions of the satellites
but not on their absolute ones in space (Section 3.2).

Jupiter's equilibrium shape deformation in the equatorial plane (Section 3.4, Figure \ref{fig:shapes}) can 
amount up to 28~m with respect to an equatorial mean radius. Wherever we would draw a dividing line, 
opposite hemispheres are always found to be asymmetric. The only  exception to that is the imposed 
north/south symmetry in our model. 
%which we know does not hold from the odd harmonics measured by Juno \citep{Kaspi18}. 

Our results lead us to conclude that taking into account the actual orbital positions of Io, Europa, and
Ganymede is important for an observational determination of Jupiter's tesseral harmonics due to tides below a 
10\% accuracy level. Even if the tesseral harmonics contribution from static tides might be too weak to be 
disentangled from the Juno gravity data, their consideration may help to reduce the uncertainty in the 
determination of other sources of longitudinal gravity field variations such as deep zonal flows \citep{Galanti17}
or the Great Red Spot \citep{Parisi16}.

In this work we the neglected dynamic effects or non-equilibrium tidal response of Jupiter. 
Methodically, this is a considerable simplification. 
On the other hand, \citet{Lainey17},  based on the method of \citet{Remus12}, computed Saturn's $k_{2}$ value from
the real part of the complex tidal Love number $k_2^c$ that does take into account tidal dissipation, for
which there is observational evidence from the orbital evolution of the Saturnian satellites. They found that
corresponding $k_{2}$ values can hardly be distinguished from the static, fluid response value. 
Further Juno measurements may illuminate us whether or not static tidal response is a valid assumption for 
estimating the shape and gravity field deformation of a gaseous planet like Jupiter.

%%%%%%%%%%%%%%%%%%%%%%%%
\acknowledgements

This work was supported by the DFG grant NE1734/1-1 of the German Science Foundation. NN thanks Daniele Durante and 
Virginia Notaro and for an inspiring discussion, Ronald Redmer for continuously helpful advice and support, Ludwig
Scheibe for fruitful feedback, and the referee for constructive comments.

%%%% Bib %%%%%%%%%%%%%%%%%%%%%%%%%%%
\bibliography{refs_knm}

%%%%%%%%%%%%%%%%%%%%%%%%%%%%%%%%%%%%%%%%%%
\appendix
%%%%%%%%%%%%%%%%%%%%%%%%%%%%%%%%%%%%%%%%%%

%%%%%%%%%%%%%%%%%%%%%%%%%%%%%%%%%%%%%%%%%%%%%%%%%%%%%%%
\section{Potential field decomposition}\label{sec:apx}

The total potential $U$ of a rotationally and tidally perturbed fluid planet can be written $U=V+Q+W$, 
where $V$ is the gravitational, $Q$ the centrifugal, and $W$ the tide-raising potential. In the following 
we describe how we  obtain the components $V_{nm}$ and $W_{nm}$ for computation of the Love numbers $k_{nm}$.

%%%%%%%%%%%%%%%%%%%%%%%%%%%%%%%%%%%%%%%%%%%%%%%%%%%%%%%%%%%%%%%%%%%%%%%%%%%
\subsection{Tide-raising potential $W$: single perturber}\label{sec:apx_W}

For a single external perturber of mass $M_S$ and orbital distance $r_S$, such as the parent star or a satellite, $W$ in a planet-centered spherical coordinate system 
$\vec{r}=(r,\varphi,\mu=\cos \theta)$ is given by $W(\vec{r}) = GM_S/|\vec{r}-\vec{r}_S|$. Using 
\begin{equation}\label{eq:taylor_rr}
	\frac{1}{|\vecrp-\vec{r}|} = \frac{1}{r}\times \sum_{n=0}^{\infty} 
	\left(\frac{r'}{r}\right)^{k} P_{n}(\cos\psi)
\end{equation}
with $k=n$ for the external field ($r>r'$), $k=-(n+1)$ for the internal field ($r<r'$), 
$\cos\psi = \vecrp \cdot \vec{r}/r'r$, and \citep{ZT78}
\begin{equation}\label{eq:PnPsi}
	P_n(\cos\psi) = P_n(\cos\theta)\,P_n(\cos\theta') +  2\sum_{m=1}^{n} \frac{(n-m)!}{(n+m)!}
	\bigg[ \left(\cos m\varphi\cos m\varphi' 
	+ \sin m\varphi\sin m\varphi'\right)P_n^m(\cos\theta)\,P_n^m(\cos\theta')\bigg]\,, \nonumber
\end{equation}
 where the $P_n$ ($P_n^m$) are the (associated) Legendre polynomials, the multipole expansion of $W$ reads 
\begin{equation} \label{eq:fullW}
	W(r,\varphi,\mu) = \frac{GM_S}{r_{\rm S}}\sum_{n=0}^{\infty} 
	\left(\frac{r}{r_S}\right)^{n} 
	\bigg[ P_n(\mu) \,P_n(\mu_S)  +\: 2\sum_{m=1}^n\frac{(n-m)!}{(n+m)!}\:
	\cos m(\varphi-\varphi_S)\:P_n^m(\mu)\,P_n^m(\mu_S)\bigg] \:.
\end{equation}
With $\cos m(\varphi-\varphi_S)=\Re\{\exp^{im(\varphi-\varphi_S)}\}$, Eq.~\ref{eq:fullW} can be written as 
\begin{equation}
	W = \sum_{n=0}^{\infty} W_n P_n(\mu) + \Re\sum _{n=0}^{\infty}\sum_{m=1}^n \,W_{nm}\, \exp^{im\varphi}\, P_n^m(\mu)
\end{equation} with
\begin{eqnarray}\label{eq:Wn}
	W_n &=& -\gamma_S\: (r/r_S)^{n}\:P_n(\mu_S)\:,\\
	\label{eq:Wnm}
	W_{nm} &=& -2 \gamma_S  \Big(\frac{r}{r_S}\Big)^{n} \frac{(n-m)!}{(n+m)!}\: \exp^{-i m\varphi_S}P_n^m(\mu_S)\:,
\end{eqnarray}
and $\gamma_S:=(1/3)\: (GM_p/R_{eq})\: q_{tid}\: (r_{S}/R_{eq})^{2}.$

%%%%%%%%%%%%%%%%%%%%%%%%%%%%%%%%%%%%%%%%%%
\subsection{$W$ of multiple perturbers}

For a number $N_{\rm sat}$ of different  tidal perturbers in the planetary equatorial plane we may assume the total tide-raising potential $W$ be given by liner superposition, 
$W(\vec{r}) = \sum_{i=1}^{N_{\rm sat}} (GM_{S_i})/|\vec{r}-\vec{r}_{S_i}|$.
Each satellite induces its own tidal forcing $q_{S_i} = -3(M_{S_i}/M_p)$ $(R_{eq}/r_{S_i})^3$. 
Equivalently to Eqs.~(\ref{eq:Wn},$\,$\ref{eq:Wnm}) we may write
\begin{equation}
	W = \sum_{n=0}^{\infty} \sum_{i=1}^{N_{\rm sat}}W_{n,i}\:P_n(\mu) 
	+ \Re \sum _{n=0}^{\infty}\sum_{m=1}^n \sum_{i=1}^{N_{\rm sat}}W_{nm,\,i}\:\exp^{im\varphi}P_n^m(\mu)
\end{equation}
with
\begin{eqnarray}
	W_{n,\,i}&=& -\gamma_{S_i}\: (r/r_{S_i})^{n}\: P_n(\mu_{S_i})\:,\\
	\label{eq:Wnmi}
	W_{nm,\,i} &=& -2 \gamma_{S_i}  \Big(\frac{r}{r_{S_i}}\Big)^{n} \frac{(n-m)!}{(n+m)!}\: 
	\exp^{-i m\varphi_{S_i}} P_n^m(\mu_{S_i}),
\end{eqnarray}
and $\gamma_{S_i}:=(1/3)\: (GM_p/R_{eq})\: q_{S_i}\: (r_{S_i}/R_{eq})^{2}$.

%%%%%%%%%%%%%%%%%%%%%%%%%%%%%%%%%%%%%%%%%%%%%%%
\subsection{Planetary gravitational field $V$}

%%% V %%%
The rotationally and tidally perturbed gravity field of a planet, $V(\vec{r})=(G/r)\int d^3r' 
\rho(\vecrp)/|\vecrp-\vec{r}|$, can be decomposed as 
$V(r)=V_0^{\rm ext} + V_0^{\rm int} + V_{rot}^{\rm ext} + V_{rot}^{int} + V_{tid}^{\rm ext} + V_{tid}^{int}$.  Notation (ext/int) refers to the contribution from mass elements interior ($r'<r$) or exterior $(r'>r)$ to the 
sphere through $\vec{r}$. Its multipole expansion can be written as
\begin{eqnarray}
	V(r,\varphi,\vartheta) &=& \frac{GM_p}{r}
	+ \frac{G}{r}\Bigg(\sum_{n=1}^{\infty} 
	\:\Bigg[\:\int_{r>r'}\hspace*{-0.3cm} d^3r'\:\rho(\vecrp) \Big(\frac{r'}{r}\Big)^{n} P_{n}(\mu') \: P_{n}(\mu)
	+ \int_{r<r'}\hspace*{-0.3cm} d^3r'\:\rho(\vecrp) \Big(\frac{r'}{r}\Big)^{\!-(n+1)} P_{n}(\mu') \: P_{n}(\mu) 
	\nonumber\\
 	&& \hspace{-0.0cm}
	+ \: 2 \sum_{m=1}^n \frac{(n-m)!}{(n+m)!} \int_{r>r'}\hspace*{-0.3cm} d^3r' \:\rho(\vecrp) 
		\Big(\frac{r'}{r}\Big)^{n} \cos m(\varphi' - \varphi)\:P_n^m(\mu') \:P_n^m(\mu) \nonumber\\
 	&& \hspace{-0.0cm}
	+ \: 2 \sum_{m=1}^n \frac{(n-m)!}{(n+m)!} \int_{r<r'}\hspace*{-0.3cm} d^3r' \:\rho(\vecrp) 
		\Big(\frac{r'}{r}\Big)^{\!\!-(n+1)} \hspace*{-0.1cm} \cos m(\varphi' - \varphi)\:P_n^m(\mu') \:P_n^m(\mu)
	\Bigg]\:\:\Bigg)\:.
\label{eq:Vmulti}
\end{eqnarray}
With the help of the abbreviations
\begin{eqnarray}
	C_{nm,\,r} &=& \frac{1}{M_p r^n}\:2\frac{(n-m)!}{(n+m)!} \int_{r>r'} d^3r' \: \rho(\vecrp) \: r'^n 
	\: \cos m\varphi' \: P_n^m(\mu')\:,\nonumber\\
	C_{nm,\,r}^{\,'} &=& \frac{r^{n+1}}{M_p}\: 2\frac{(n-m)!}{(n+m)!}\int_{r<r'}d^3r' \:\rho(\vecrp)\:
	\frac{\cos m\varphi'}{r'^{(n+1)}} \: P_n^m(\mu')\:,\nonumber\\
	S_{nm,\,r} &=& \frac{1}{M_p r^n}\: 2\frac{(n-m)!}{(n+m)!}\int_{r>r'} d^3r' \: \rho(\vecrp) \: r'^n 
	\: \sin m\varphi' \: P_n^m(\mu')\:, \nonumber\\
	S_{nm,\,r}^{\,'} &=& \frac{r^{n+1}}{M_p}\: 2\frac{(n-m)!}{(n+m)!}\int_{r<r'}d^3r' \:\rho(\vecrp)\:
	\frac{\sin m\varphi'}{r'^{(n+1)}}  \: P_n^m(\mu')\:,\nonumber\\
 &&
\end{eqnarray}
and $C_{nm}\cos m\varphi=\Re\,C_{nm}\exp^{im\varphi}$, $S_{nm}\sin m\varphi$ $=\Re\{ -i S_{nm}$$\exp^{im\varphi}\}$,
Eq.~(\ref{eq:Vmulti}) can be written 
\begin{equation}	\label{eq:VVnVnm}
	V(r,\varphi,\mu) = \frac{GM_p}{r} \:+\: \sum_{n=1}^{\infty} (V_n^{ext} + V_n^{int})\:  P_n(\mu)
	 \:+\: \Re\:\sum_{n=1}^{\infty}\sum_{m=1}^{n} (V_{nm}^{ext} + V_{nm}^{int})\exp^{im\varphi}\:P_n^m(\mu)
\end{equation}
with 
$V_{nm}^{ext} = (GM_p/r)\: (C_{nm,\,r} - i\, S_{nm,\,r})$,
$V_{nm}^{int} = (GM_p/r)\: (C_{nm,\,r}^{\,'} - i\, S_{nm,\,r}^{\,'})\:$.
In Equation (\ref{eq:VVnVnm}), the first term equals the total gravitational potential of the unperturbed planet 
$V_0^{ext}+V_0^{int}$, the second term describes the rotational potential $V_{rot}$, and the third term can be identified
as the tidal perturbation $V_{tid}$.

%%%%%%%%%%%%%%%%%%%%%%%%%%%%%%%%%%%%%%%%%%%%%%%%%%%%%
\subsection{Love numbers $k_{nm}$}\label{sec:apx_knm}

We assume real-valued and linear tidal response coefficients $k_{nm}$. They are obtained by from the definition 
$V_{nm} = k_{nm}\: W_{nm}$ evaluated on the surface of the planet $r=R(\varphi,\mu)$. There, the $V^{int}_{nm}$ vanish
and we denote $V_{nm}=V^{ext}_{nm}(R)$.  
For a point on the equator  we  have $R=R_{eq}$ and thus
\begin{equation}
	\frac{GM_p}{R_{eq}}\:(C_{nm} - i\, S_{nm}) = - k_{nm}\: \frac{2}{3}\: \frac{GM_p}{R_{eq}}\: 
	\Bigg( \sum_{i=1}^{N_{\rm sat}} q_{S_i} \times
	\Big(\frac{r_{S_i}}{R_{eq}}\Big)^{2}	\Big(\frac{R_{eq}}{r_{S_i}}\Big)^{n} \frac{(n-m)!}{(n+m)!}\: 
	\exp^{-i m\varphi_{S_i}} P_n^m(\mu_{S_i})\Bigg)
\end{equation}
The static and immediate tidal response in phase with the tidal forcing is associated 
with the real part of the Love numbers $k_{nm}$,
\begin{equation}\label{eq:Ree_knm}
	\Ree\:k_{nm}=-\frac{3}{2}\:\frac{(n+m)!}{(n-m)!}\:\Ree \left\{ 
	\frac{C_{nm} - i\,S_{nm}}{\sum_{i=1}^{N_{\rm sat}} q_{S_i} \Big(\frac{R_{eq}}{r_{S_i}}\Big)^{n-2} 
	P_n^m(\mu_{S_i}) \exp^{-i m\varphi_{S_i}} }
 \right\}\:,
\end{equation}
which we may abbreviate as $k_{nm} = A_{nm}/B_{nm}$,
while all other effects including dynamics, dissipation, and out-of-phase response is associated with the 
imaginary part of the $k_{nm}$ \citep{Ogilvie14}. Equation (\ref{eq:Ree_knm}) directly yields Eq.~(\ref{eq:knm_multi1}). 
Here we request $\Imm\: k_{nm}=0$, in which case $\Ree k_{nm} = \Ree A_{nm}/\Ree B_{nm}$ and thus we obtain
Eq.~(\ref{eq:knm_multi2}).

\end{document}